\documentclass[preprint,prd,tightenlines,superscriptaddress]{revtex4-1}

\usepackage{bm,amsmath,amsfonts,amssymb,graphicx,xspace,placeins,multirow}

\usepackage{graphicx} 
\usepackage{dcolumn}  
\usepackage{colordvi}
\usepackage{color}
\usepackage{epstopdf}
\usepackage{amssymb}
\usepackage{url}
\graphicspath{{ps}}
\usepackage{hyperref}
\usepackage{tabularx}
\usepackage{lineno}
\usepackage{adjustbox}
\usepackage{hyperref}

\newcommand{\bdslnu}{\ensuremath{B \to D^{*} \ell \nu_\ell}\xspace}

\newcommand{\bdlnu}{\ensuremath{B \to D \ell\nu_\ell}\xspace}

\newcommand{\bchdenu}{\ensuremath{\Bp \to \bar D^{0} e^{+} \nu_e}\xspace}
\newcommand{\bchdmunu}{\ensuremath{\Bp \to \bar D^{0} \mu^{+} \nu_\mu}\xspace}
\newcommand{\bneudenu}{\ensuremath{\Bz \to D^{-} e^{+} \nu_e}\xspace}
\newcommand{\bneudmunu}{\ensuremath{\Bz \to D^{-} \mu^{+} \nu_\mu}\xspace}

\newcommand{\bneudellnu}{\ensuremath{\Bz \to D^{-} \ell^{+} \nu_\ell}\xspace}
\newcommand{\bchdellnu}{\ensuremath{\Bp \to \bar D^{0} \ell^{+} \nu_\ell}\xspace}

\newcommand{\ifb}{\ensuremath{{\rm fb}^{-1}}\xspace}
\newcommand{\lumi}{\ensuremath{189.2~\ifb}\xspace}
\newcommand{\BR}{{\ensuremath{\cal B}}}

\newcommand{\cosby}{\ensuremath{\cos\theta_{BY}}}

\newcommand{\resVcb}{\ensuremath{\eta_\mathrm{EW}|V_{cb}|=(38.53\pm 1.15)\times 10^{-3}}\xspace}

\newcommand{\resVcbnoeta}{\ensuremath{|V_{cb}|=(38.28\pm 1.16)\times 10^{-3}}\xspace}
\newcommand{\NBB}{\ensuremath{N_{B \bar B}=(198.0\pm 3.0)\times 10^6}}

\newcommand{\dGidw}{\ensuremath{\Delta \Gamma_i / \Delta w}}

\newcommand{\dGiBGLdw}{\ensuremath{\Delta \Gamma_{i,\mathrm{BGL}} / \Delta w}}

\newcommand{\dGidwf}{\ensuremath{\frac{\Delta \Gamma_i}{ \Delta w}}}
\newcommand{\dGjdwf}{\ensuremath{\frac{\Delta \Gamma_j}{ \Delta w}}}

\newcommand{\dGiBGLdwf}{\ensuremath{\frac{\Delta \Gamma_{i,\mathrm{BGL}}}{ \Delta w}}}
\newcommand{\dGjBGLdwf}{\ensuremath{\frac{\Delta \Gamma_{j,\mathrm{BGL}}}{ \Delta w}}}

\input ./belle2sym.tex

\begin{document}

\def\belletwo {\it {Belle II}}

\vspace*{-3\baselineskip}
\resizebox{!}{3cm}{\includegraphics{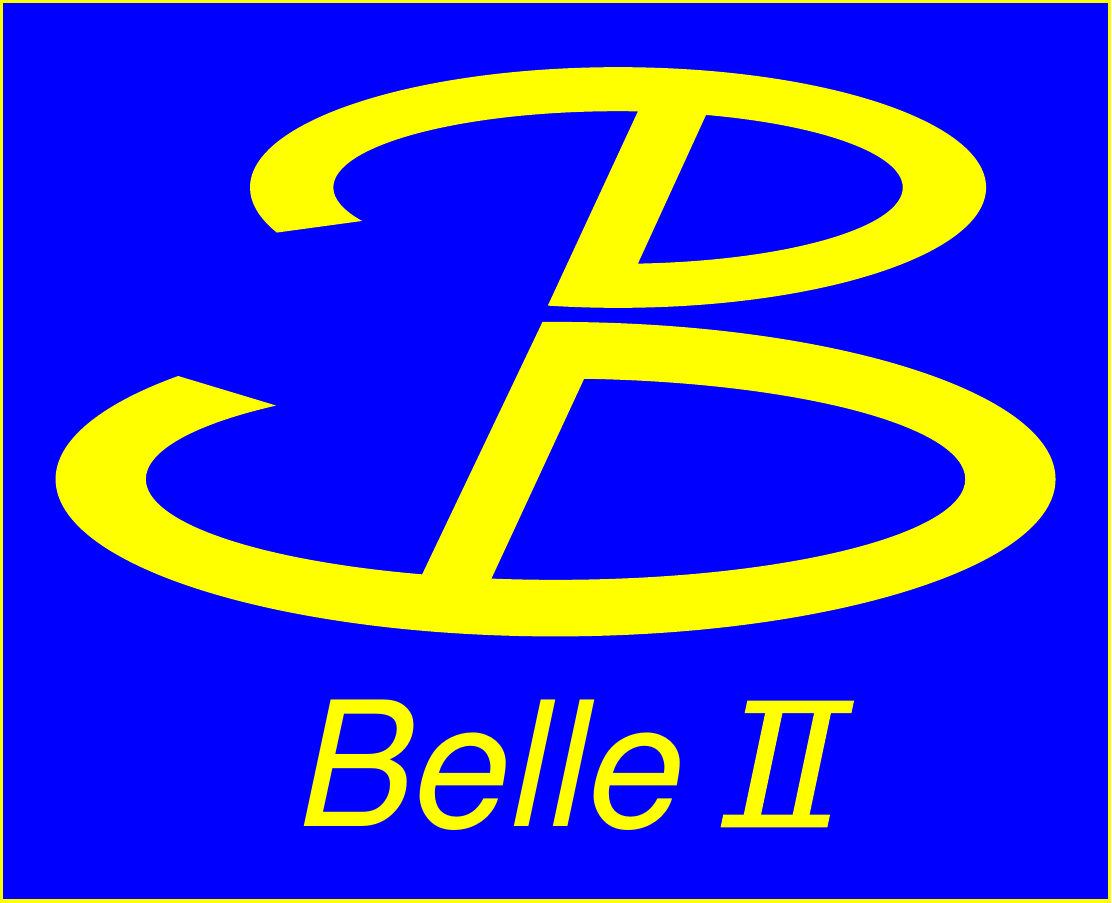}}

\vspace*{-5\baselineskip}
    \begin{flushright}
BELLE2-CONF-PH-2022-010 \\
\today
\end{flushright}

\title { \quad\\[0.5cm] Determination of $|V_{cb}|$ from $B\to D\ell\nu$ decays using 2019--2021 Belle II data }

\collaboration{The Belle II Collaboration}
  \author{F. Abudin{\'e}n}
  \author{I. Adachi}
  \author{K. Adamczyk}
  \author{L. Aggarwal}
  \author{P. Ahlburg}
  \author{H. Ahmed}
  \author{J. K. Ahn}
  \author{H. Aihara}
  \author{N. Akopov}
  \author{A. Aloisio}
% ...

\begin{abstract}
We present a determination of the magnitude of the Cabibbo-Kobayashi-Maskawa (CKM) matrix element $V_{cb}$ using $B\to D\ell\nu$ decays. The result is based on $e^+e^-\to\Upsilon(4S)$~data recorded by the Belle II detector corresponding to \lumi of integrated luminosity. The semileptonic decays \ensuremath{\Bz \to D^{-}(\to K^+ \pi^- \pi^-) \ell^{+} \nu_\ell}\xspace and \ensuremath{\Bp \to \bar D^{0}(\to K^+\pi^-) \ell^{+} \nu_\ell}\xspace are reconstructed, where $\ell$ is either electron or a muon. The second $B$~meson in the $\Upsilon(4S)$~event is not explicitly reconstructed. Using the diamond-frame method, we determine the $B$~meson four-momentum and thus the hadronic recoil. We extract the partial decay rates as functions of $w$ and perform a fit to the decay form-factor and the CKM parameter $|V_{cb}|$ using the BGL parameterization of the form factor and lattice QCD input from the FNAL/MILC and HPQCD collaborations. We obtain \resVcb, where $\eta_{EW}$ is an electroweak correction, and the error accounts for theoretical and experimental sources of uncertainty.

\end{abstract}


\maketitle

{\renewcommand{\thefootnote}{\fnsymbol{footnote}}}
\setcounter{footnote}{0}

\section{Introduction}

The magnitude of the Cabibbo-Kobayashi-Maskawa (CKM) matrix element $V_{cb}$ squared determines the transition rate of $b$ into $c$ quarks~\cite{Cabibbo:1963yz,Kobayashi:1973fv}. Precise knowledge of this fundamental parameter of the Standard Model (SM) is crucial for the ongoing precision-$B$-physics program at the Belle II experiment and elsewhere. The CKM magnitude $|V_{cb}|$ is measured from semileptonic $B\to X_c\ell\nu$ decays, where $B$ is either $B^+/B^-$ or $B^0 /\bar B^0$, $X_c$ is a hadronic system with a charm quark, $\ell$ is a light charged lepton (electron or muon), and $\nu$ is the associated neutrino. These determinations can be $\emph{inclusive}$, {\it i.e.}, based on all $X_c\ell\nu$ final states within a given region of phase space, or $\emph{exclusive}$, {\it i.e.}, based only on a single $b\to c$ semileptonic decay mode such as $B\to D^*\ell\nu$ or $B\to D\ell\nu$. Pursuing both approaches is important as they involve different theoretical and experimental uncertainties and consistency is a powerful cross-check. However, inclusive and exclusive measurements of $|V_{cb}|$ have persistently shown an approximate $3.3 \sigma$ discrepancy \cite{HFLAV:2019otj}.

This paper describes a measurement of the decay $B\to D\ell\nu$ (\bneudellnu and \bchdellnu) in $\Upsilon(4S)$~events and a determination of $|V_{cb}|$ based on Belle II data corresponding to \lumi of integrated luminosity~\cite{cc}. To maximize the statistical power of the early Belle II data set, the sample is untagged, {\it i.e.}, we do not explicitly reconstruct the second $B$~meson in the $\Upsilon(4S)$ event. The disadvantage of this approach are large combinatorial backgrounds from other semileptonic modes, especially $B \to D^* \ell \nu$.
Throughout the text we refer to \bneudellnu as the \textit{neutral mode}, and \bchdellnu as the \textit{charged mode} in reference to the $B$ meson charge.  The document is organized as follows: Section~\ref{sec:theo} introduces the theory of the $B\to D\ell\nu$ decay and of the measurement of $|V_{cb}|$, and Sect.~\ref{sec:exp} describes our experimental procedure. Section~\ref{sec:res} contains the results of this analysis, the partial branching fractions as a function of $w=v\cdot v'$ along with a discussion of the systematic uncertainty. Finally, in Sect.~\ref{sec:vcb} we report the final results for the decay form-factors and $|V_{cb}|$.

\section{Theoretical framework} \label{sec:theo}

The differential decay rate of $B\to D\ell\nu$ is usually expressed as a function of $w=v\cdot v'$, where $v$ and $v'$ are the four-velocities of the $B$ and the $D$~mesons, respectively. The quantity $w$ is related to $q^2$, the four-momentum squared of the lepton neutrino system by
\begin{equation}
    w=\frac{m_B^2+m_D^2-q^2}{2m_B m_D}~\label{eq:w},
\end{equation}
 where $m_B$ and $m_D$ are the $B$ and $D$ meson masses.
The range of $w$ is bounded by the zero recoil point ($w=1$), where the $D$~meson is at rest in the $B$~frame and by
\begin{equation}
      w_\mathrm{max} = \frac{m_B^2+m_D^2}{2m_Bm_D} \approx 1.59~,
\end{equation}
where the entire $B$~energy is transferred to the $D$~meson. Neglecting the lepton mass, the expression for the differential decay rate reads~\cite{Neubert:1993mb}
\begin{equation}
  \frac{d \Gamma(B\to D\ell\nu_\ell)}{d w} = \frac{G^2_\mathrm{F} m^3_D}{48\pi^3}(m_B+m_D)^2(w^2-1)^{3/2}\eta_\mathrm{EW}^2\mathcal{G}^2(w)|V_{cb}|^2~, \label{eq:rate}
\end{equation}
where $G_\mathrm{F}$ is Fermi's constant and $\eta_\mathrm{EW}$ is an electroweak correction~\cite{Sirlin:1981ie}. The form factor $\mathcal{G}(w)$ contains the single amplitude $f_+(w)$,
\begin{equation}
  \mathcal{G}^2(w) = \frac{4r}{(1+r)^2} f^2_+(w)~, \label{eq:ff}
\end{equation}
where $r=m_D/m_B$.

Various parameterizations of the form factor $\mathcal{G}(w)$ are available. The most commonly used expression is the one of Caprini, Lellouch, and Neubert (CLN)~\cite{Caprini:1997mu}. It reduces the number of free parameters by adding multiple dispersive constraints based on spin- and heavy-quark symmetries,
\begin{equation}
  \mathcal{G}(z)= \mathcal{G}(1)\big(1 - 8 \rho^2 z + (51 \rho^2 - 10 )
  z^2 - (252 \rho^2 - 84 ) z^3\big)~,
  \label{eq:CLN}
\end{equation}
where
\begin{equation}
  z(w) = \frac{\sqrt{w + 1} - \sqrt{2}  }{\sqrt{w + 1} + \sqrt{2}  }~.
\end{equation}
The only two free parameters are the form factor at zero recoil $\mathcal{G}(1)$ and the linear slope $\rho^2$. The precision of this approximation 
is estimated to be better than 2\%, which is close to the current experimental accuracy of $|V_{cb}|$.

As a consequence, a model-independent expression that relies only on QCD dispersion relations has been proposed by Boyd, Grinstein, and Lebed (BGL)~\cite{Boyd:1994tt},
\begin{equation}
 f_i(z) = \frac{1}{P_i(z) \phi_i(z)} \sum\limits_{n=0}^{N} a_{i,n} z^n, \quad i=+,0 \label{eq:BGL}~,
\end{equation}
where $P_i(z)$ are the {\it Blaschke factors}, containing explicit poles ({\it e.g.}, the $B_c$ or $B^*_c$ poles) in $q^2$ and $\phi_i(z)$ are the 
{\it outer functions}, which are arbitrary but required to be analytic without any poles or branch cuts. The $a_{i,n}$ coefficients are free parameters and $N$ is 
the order at which the series is truncated. Following Ref.~\cite{Lattice:2015rga}, we choose $P_i(z)=1$ and
\begin{eqnarray}
\phi_+(z) & = & 1.1213 (1+z)^2 (1-z)^{1/2}
           [(1+r)(1-z)+ 2\sqrt{r}(1+z)]^{-5}~, \label{eq:phiplus}\\
\phi_0(z) & = & 0.5299 (1+z)(1-z)^{3/2}  
	   [(1+r)(1-z)+ 2\sqrt{r}(1+z)]^{-4}~. \label{eq:phizero}
\end{eqnarray}
The parameterization also contains the form factor $f_0$, which is relevant only in semitauonic decays. Theoretical calculations are also available for $f_0$ and can provide constraints through the kinematic constraint at maximum recoil $w_\mathrm{max}\approx 1.6$,
\begin{equation}
  f_0(w_\mathrm{max}) = f_+(w_\mathrm{max})~. \label{eq:kinematic}
\end{equation}

The procedure for obtaining the CKM matrix element $|V_{cb}|$ depends on the form factor used: for CLN, the differential rate is fit to Eqs.~\ref{eq:rate} and \ref{eq:CLN} and $\eta_\mathrm{EW}\mathcal{G}(1)|V_{cb}|$ and $\rho^2$ are obtained. To determine $|V_{cb}|$, theory input for $\mathcal{G}(1)$ is required. However, in Sect.~\ref{sec:vcb} we will employ a method based on the BGL parameterization. This involves a combined fit to our differential rate in intervals (\textit{bins}) of $w$ (Sect.~\ref{sec:res}) and lattice QCD calculations of $f_i(w)$ ($i=+,0$)~\cite{Lattice:2015rga,Na:2015kha} to Eqs.~\ref{eq:rate}, \ref{eq:ff}, \ref{eq:BGL} and \ref{eq:kinematic} for a fixed order $N$. The results of the fit are $\eta_\mathrm{EW}|V_{cb}|$ and the coefficients of the BGL expansion $a_{i,n}$.

\section{Experimental procedure} \label{sec:exp}

\subsection{Data sample and event selection}

The Belle~II detector~\cite{Abe:2010sj} operates at the SuperKEKB asymmetric-energy  electron-positron collider~\cite{superkekb}, located at KEK in Tsukuba, Japan. The detector consists of several nested detector subsystems arranged around the beam pipe in a cylindrical geometry. The innermost subsystem is the vertex detector, which includes two layers of silicon pixel detectors and four outer layers of silicon strip detectors. Currently, the second pixel layer is instrumented in only a small part of the solid angle. Most of the tracking volume consists of a helium and ethane-based small-cell drift chamber (CDC). Outside the drift chamber, a Cherenkov-light imaging and time-of-propagation detector (TOP) provides charged-particle identification in the barrel region. In the forward endcap, this function is provided by a proximity-focusing, ring-imaging Cherenkov detector with an aerogel radiator (ARICH). Further out is the ECL electromagnetic calorimeter, consisting of a barrel and two endcap sections made of CsI(Tl) crystals. A uniform 1.5~T magnetic field is provided by a superconducting solenoid situated outside the calorimeter. Multiple layers of scintillators and resistive-plate chambers, located between the magnetic flux-return iron plates, constitute the $K^0_L$ and muon identification system (KLM).

The data used in this analysis were collected in the years 2019 to 2021 at a center-of-mass (c.m.) energy of 10.58~GeV, corresponding to the mass of the $\Upsilon$(4S) resonance.
This data set corresponds to an integrated luminosity of \lumi and contains \NBB\ $\Upsilon(4S)\to B\bar B$ events as determined from a fit to event-shape variables. In addition, we use a sample corresponding to 18 fb$^{-1}$ of off-resonance collision data, collected at a c.m.\ energy 60 MeV below the $\Upsilon(4S)$ resonance.

Two samples of Monte-Carlo-simulated (MC) events are used. These are a sample of $\Upsilon(4S)\to B\bar B$~events in which $B$~mesons decay generically, generated with EvtGen~\cite{Lange:2001uf}, and a sample of continuum $e^+e^-\to q\bar q$~events ($q = u, d, s, c$) simulated with KKMC~\cite{Ward:2002qq} interfaced with PYTHIA~\cite{Sjostrand:2007gs}. The simulation of semileptonic $B$~decays includes so-called \emph{gap modes} ($B\to D^{(*)}\pi\pi\ell\nu$ and $B\to D^{(*)}\eta\ell\nu$) to account for the discrepancy between the sum of known exclusive semileptonic branching fractions and the inclusive $B$~decay rate~\cite{zyla:2020}. Full detector simulation based on GEANT4~\cite{ref:geant4} is applied to MC~events. The Monte Carlo samples used in this analysis correspond to an integrated luminosity of about 1~ab$^{-1}$. The lepton reconstruction efficiencies and the hadron misidentification rates in simulation are adjusted to match the performance of the Belle II lepton identification system in data.
The data samples are processed using the Belle II software framework basf2~\cite{Kuhr:2018lps}.

Prior to physics analysis, charged-particle trajectories ({\it tracks}) are reconstructed in the vertex detector and the CDC~\cite{Bertacchi:2020eez}. Photons are reconstructed from localized energy deposits in the ECL ({\it clusters}) unmatched to tracks. Hadronic events are selected by requiring at least three charged particles in the $\Upsilon(4S)$~event, and a ratio $R_2$ of the second to the zeroth Fox-Wolfram moment below 0.4~\cite{Fox:1978vu}. The visible energy, i.e., the sum of energies associated to all particle tracks and clusters observed in the event, is required to be above 4~GeV in the c.m.\ frame. For improved suppression of continuum background, we also require the Kakuno-Super-Fox-Wolfram (KSFW) moment  $H^{so}_{20}$~\cite{BaBar:2014omp} to be greater than 0.18.

\subsection{Signal reconstruction}

Tracks are required to originate from the interaction point by imposing a distance of closest approach of less than 3 cm along the $z$ direction (parallel to the beams) and less than 1~cm in the transverse plane. We further require charged particles to be within the CDC angular acceptance and have a sufficient number of associated CDC hits. In the following all quantities are defined in the laboratory frame unless otherwise stated.

Electron and muon candidates are identified using particle-identification information and have a c.m.\ momentum $p^*_\ell$ greater than 0.6~GeV. Electrons are identified mainly by comparing the energy measured in the ECL and the momentum measured using tracking. Muons are identified mainly using information from the instrumented return yoke or KLM. We partially recover bremsstahlung photons radiated from an electron by searching within a cone around the lepton direction. For electron momenta $p_e < 1$~GeV,  the photon is required to have $E<0.9$~GeV and the angle between the photon and the electron direction has to be smaller than 0.137~rad. For electrons above 1~GeV, we require $E < 1.2$~GeV and an angle smaller than $0.074$~rad. If such photons are found their four-momenta are added to the electron-candidate.

Kaons are identified by combining information from the TOP, ARICH, and CDC. Their momentum is required to be larger than $0.5$ GeV. Candidate $D$~mesons are searched for in the decay modes $D^0\to K^-\pi^+$ and $D^+\to K^- \pi^+ \pi^+$. For $D^0$~candidates, an identified $K$~candidate is combined with a oppositely charged $\pi$ and the $K\pi$~invariant mass is required to lie within the 1.85~GeV to 1.88~GeV range. For $D^+$~candidates we apply the selection $1.86<M(K\pi\pi)< 1.88$~GeV. In both cases the mass ranges correspond to about three times the  mass resolution of each mode. 

Candidate $B\to D\ell\nu$ decays are formed by combining an appropriately charged lepton with a $D$~candidate. The mass of the $Y = D \ell$ system is required to exceed 3 GeV. For each $B$~candidate, we calculate the following observable,
\begin{equation}
\cos\theta_{BY} = {2\, E_B^* E_Y^* - m_B^2 - m_Y^2 \over 2 p_B^* p_Y^*},\label{eq:costheta}
\end{equation}
where $E_Y^*$, $|p_Y^*|$, and $m_Y$ are the c.m.\ energy, momentum, and invariant mass, respectively, of the $D \ell$ system; $m_B$ is the known $B$ mass~\cite{zyla:2020}; and $E_B^*$ and $|p_B^*|$ are the magnitudes of the c.m.\ energy and momentum, respectively, of the $B$ candidate. The latter are inferred from the $\Upsilon(4S)$ c.m.\ energy. For correctly reconstructed $B\to D\ell\nu$~candidates, $\cos\theta_{BY}$ corresponds to the angle between the $Y$ and the $B$ meson in the c.m.\ frame and lies in the range $[-1,1]$. However, due to the finite beam-energy spread, final-state radiation, and detector resolution, the $\cos\theta_{BY}$ distributions of signal events are smeared beyond this range. For background candidates, $\cos\theta_{BY}$ values outside of the range $[-1,1]$ are allowed. In the rest of the analysis, we retain $B$~candidates with a value of $\cos\theta_{BY}$ ranging between $-4$ and 4.

Particles in the event not used in the reconstruction of the signal candidate are assigned to the rest-of-event system (ROE) and selections are applied to further improve signal purity. ROE charged particles are required to have c.m.\ momentum $p^* < 3.2$~GeV. For photons the weighted number of crystals in a cluster must be below 1.5, the signal must be in time with the collision and the cluster must lie within CDC acceptance. In addition, we require the transverse momentum corresponding to the energy detected in each  cluster to exceed 0.06 GeV to reduce the contribution of beam-related backgrounds and electronic noise. We require the ROE invariant mass to be $M_{\mathrm{ROE}} < 6$~GeV for the charged mode and $M_{\mathrm{ROE}} < 5.2$~GeV for the neutral mode. The total ROE momentum is required to be smaller than $2.8$~GeV. In addition, the missing momentum, defined as the difference between the momentum of the colliding particles and the vector sum of momenta of all charged particles and of momenta corresponding to the neutral ECL clusters, is required to be greater than 1.2~GeV for both modes.

To reduce the sizeable background from $B^0\to D^{*-}(\bar D^0\pi^-)\ell^+\nu_\ell$~decays in the charged $B$ mode, a veto is applied. This is done by combining a low-momentum (slow) pion ($p<0.35$~GeV) with the $\bar D^0$ from a $B^+\to\bar D^0\ell^+\nu_\ell$ candidate. If the mass difference $\Delta M \equiv M(K^+\pi^-\pi^-) - M(K^+\pi^-)$ is found to be in the interval $[0.144,0.148]$~GeV, the $B^+$~candidate is rejected.

Finally, a vertex fit is performed on the full decay chain using the {\tt TreeFitter} package~\cite{Belle-IIanalysissoftwareGroup:2019dlq}. 

\subsection{Reconstruction of the kinematic variable $w$} \label{sec:w}

The $B\to D\ell\nu$ form factor depends on the single kinematic variable $w$, Eq.~\eqref{eq:w}, which parameterizes the recoil momentum of the $D$~meson. To reconstruct $w$, we use the {\it diamond-frame} approach \cite{BaBar:2006taf}. The $B$ three-momentum lies on a cone around the $Y=D\ell$ direction defined by Eq.~\ref{eq:costheta}. We calculate the $B$~momentum for four uniformly distributed positions on this cone, defined by the $D$ and $\ell$ directions. For each $p_B$ solution, the kinematic variable $w$ is calculated and a weighted average of these four solutions is taken as our estimate of $w$. Taking into account the kinematics of the decay, the assigned weights are $\sin^2(\theta_B)$ where $\theta_B$ is the azimuthal angle of the $B$~meson with respect to the $D\ell$~plane measured in the center-of-mass frame.

\begin{figure}
    \centering
    \includegraphics[width=.44\columnwidth]{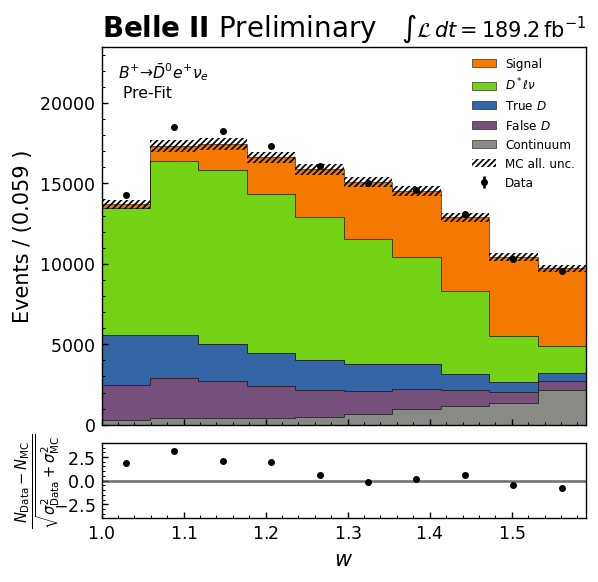}
    \includegraphics[width=.44\columnwidth]{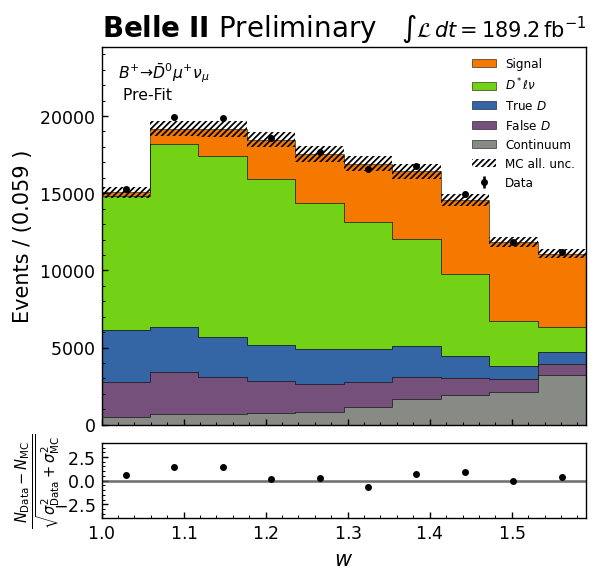}
    \includegraphics[width=.44\columnwidth]{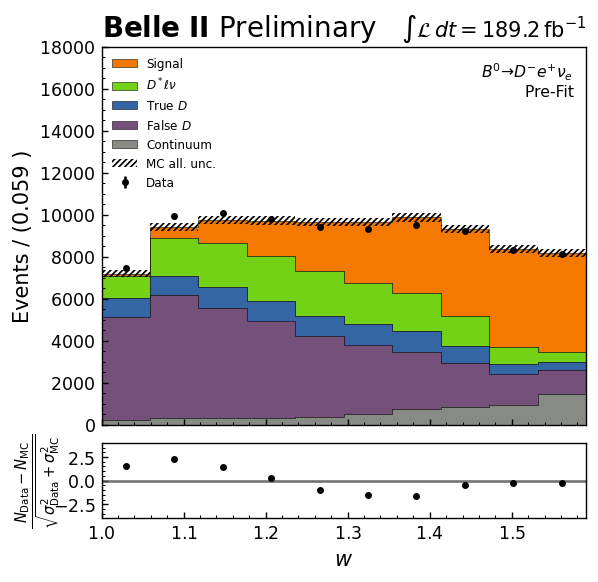}
    \includegraphics[width=.44\columnwidth]{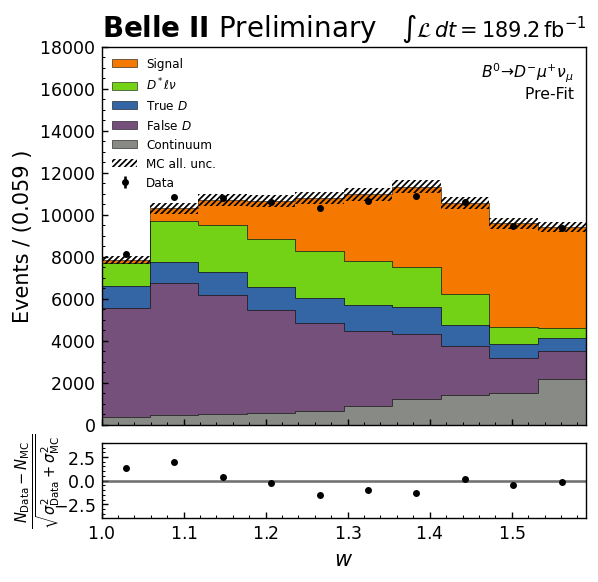}
    \caption{Reconstructed $w$~distribution, shown separately for the charged and neutral channels and the electron and muon samples. The Belle II data are the points with error bars. The stacked histograms are simulated events normalized to the data luminosity. The panels at the bottom of each distribution show the difference between data and fit results, divided by statistical and systematic standard deviations summed in quadrature.} \label{fig:w_prefit}
\end{figure}

The resulting $w$ resolution is estimated with a Gaussian fit to be 0.026 for both the charged and the neutral channels. Figure~\ref{fig:w_prefit} shows the reconstructed $w$~distribution, separately for the charged and neutral channels and the electron and muon samples.

\subsection{Signal extraction}

The signal yield in the reconstructed samples is extracted by performing a fit to the $\cos\theta_{BY}$~distribution. We use a binned maximum likelihood fit assuming Poisson statistics for both experimental data and MC simulation~\cite{Barlow:1993dm} and consider four background components in addition to the signal component: feed-down background from the decay $B\to D^*\ell\nu$, candidates with a correctly reconstructed $D$~meson in which the $B$~meson did not decay into $D^{*}\ell\nu$ (so-called "true $D$'s"), and candidates in which the $D$~meson is misreconstructed ("fake $D$'s"). Finally, non-$B \bar B$ candidates from processes such as $e^+ e^-\to q\bar q~ (u\bar u,~ d\bar d,~ s \bar s,~ \text{and }  c \bar c)$ and $e^+ e^- \to \tau^+ \tau^-$ are combined into the continuum background category. The fit is performed separately in 10~bins of $w$ ranging from 1 to $w_\mathrm{max}\approx 1.59$. The analysis is also done separately in the charged and neutral channels, and in the electron and muon samples.

Free parameters are the normalizations of the signal, $D^*\ell\nu$~feed-down, true $D$ and fake $D$ components. The probability density function (PDF) of the continuum component is taken from continuum simulation while the normalization is fixed to the level of continuum estimated using off-resonance data. We confirm that the continuum shape is consistent with the data recorded below the $\Upsilon(4S)$ resonance. In the \bchdenu and \bchdmunu modes, where $D^*\ell\nu$~feed-down is particularly important, this component is fixed to the known~value of the $B\to D^*\ell\nu$ branching fraction~\cite{zyla:2020}.

We perform the fit for values of $\cos\theta_{BY}$ ranging between $-4$ and $2$. We confirm the stability of the fit result with respect to the upper boundary in $\cos\theta_{BY}$. Simplified simulated experiments drawn from the likelihood (toy MC) show that the estimator is unbiased and has proper uncertainties.

\begin{figure}
    \centering
    \includegraphics[width=.47\columnwidth]{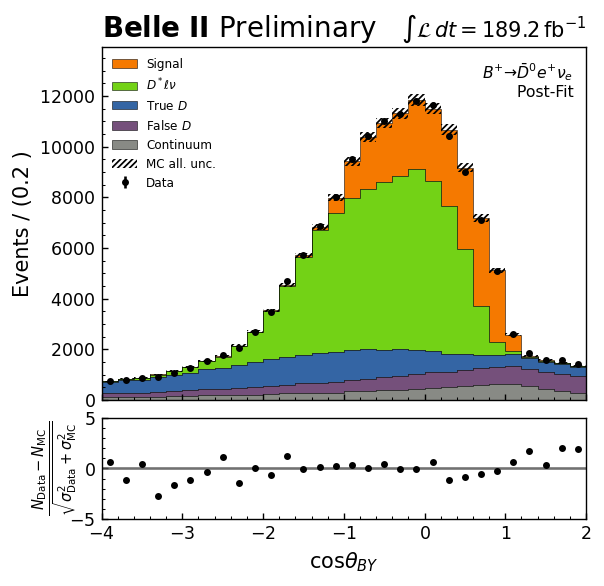}
    \includegraphics[width=.47\columnwidth]{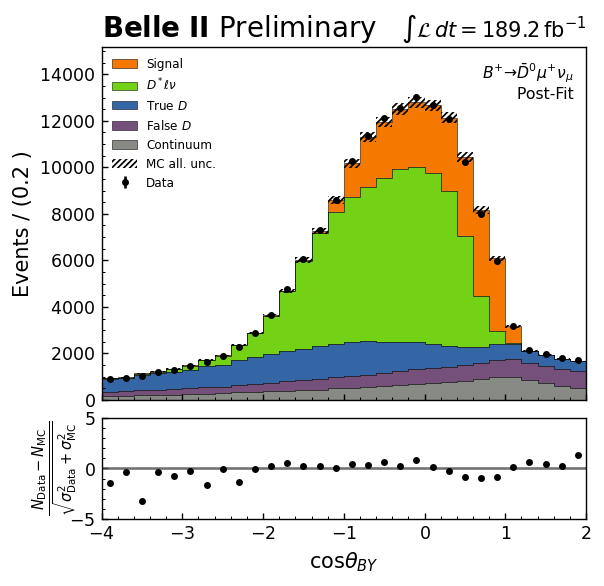}
    \includegraphics[width=.47\columnwidth]{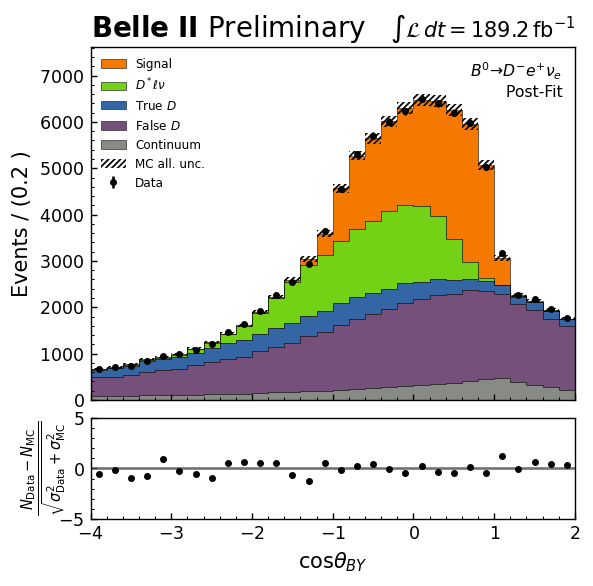}
    \includegraphics[width=.47\columnwidth]{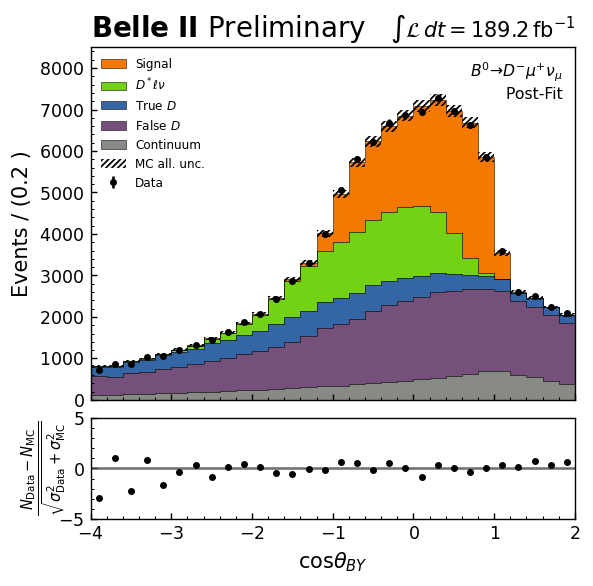}
    \caption{Distributions of $\cos\theta_{BY}$ for the four samples, with fit results overlaid. The points with error bars correspond to the Belle II data. The stacked histograms are simulated events scaled to match the result of the fit.} \label{fig:postfit}
\end{figure}

The results of the fit are shown in Table~\ref{tab:yields} and Fig.~\ref{fig:postfit}, separately for the charged and neutral channels and the electron and muon samples integrated over $w$.
\begin{table}[]
    \centering
    \begin{adjustbox}{center}
    \begin{tabular}{ccccc}
    \hline\hline\\[-2.0ex]
Component& \bchdenu& \bchdmunu& \bneudenu& \bneudmunu\\[0.5ex]\hline\\[-2.0ex]
Signal& $ 27485 \pm  388  $ & $ 29015 \pm  402  $ & $ 22824 \pm  464  $ & $ 24658 \pm  478  $ \\
$D^* \ell \nu$& $ 71761 $& $ 76808 $& $ 17256  \pm  530$& $ 17550  \pm  554$\\
True $D$& $ 22790  \pm  551$& $ 27757  \pm  593$& $ 9140  \pm  711$& $ 12607  \pm  741$\\
Fake $D$& $ 12898  \pm  516$& $ 14674  \pm  563$& $ 36702  \pm  768$& $ 39658  \pm  793$\\
Continuum& $ 9529 $& $ 15285 $& $ 6784 $& $ 10938 $ \\[0.5ex]\hline
Data yield& $ 148769 $& $ 165076 $& $ 92821 $& $ 103960$\\[0.5ex]
\hline \hline
    \end{tabular}
    \end{adjustbox}
    \caption{Fit results in the four $B\to D\ell\nu$~samples. The observed total event yield in data is displayed along with the individual fit components. The uncertainties only account for the statistical contributions. Results with no uncertainty refer to components fixed in the fit.}
    \label{tab:yields}
\end{table}

\section{Results and systematic uncertainty} \label{sec:res}

\subsection{$B\to D\ell\nu$ branching fraction}

The fit result integrated over $w$ (Table~\ref{tab:yields}) is converted into a measurement of the $B\to D\ell\nu$~branching ratios using
\begin{equation}
    N_\mathrm{sig}=2 N_{B\bar B} f_i {\cal B}(B\to D\ell\nu) {\cal B}(D)\epsilon~,
\end{equation}
where $N_{B\bar B}$ is the number of $\Upsilon(4S)$~events in the sample, $f_i$ ($i={+-}, {00}$) are the $B^+ B^-$, $B^0 \bar B^0$~production fractions at the $\Upsilon(4S)$~\cite{zyla:2020}, ${\cal B}(D)$ is the $D$~decay branching fraction~\cite{zyla:2020}, and $\epsilon$ is the total efficiency including acceptance. The results obtained in the four samples and the various contributions to systematic uncertainty are shown in Table~\ref{tab:br}.
\begin{table}
    \centering
    \begin{adjustbox}{center}
    \begin{tabular}{lcccc}
    \hline\hline\\[-2.0ex]
& \bchdenu& \bchdmunu& \bneudenu& \bneudmunu\\[0.5ex] \hline\\[-2.0ex]
$\mathcal{B}(B \rightarrow D \ell \nu)[\%] $& $ 2.21 \pm  0.03 \pm  0.08$& $ 2.22 \pm  0.03 \pm  0.10$& $ 1.99 \pm  0.04 \pm  0.08$& $ 2.03 \pm  0.04 \pm  0.09$\\[0.5ex]
\hline
  & \multicolumn{4}{c}{Contributions to the systematic uncertainty $[\%]$}\\ 
$N_{BB}$ and $f_{+-}/f_{00}$ &  1.9&  1.9&  1.9&  1.9\\
Tracking efficiency &  0.9&  0.9&  1.2&  1.2\\
$\mathcal{B}(D \rightarrow K \pi (\pi))$ &  0.8&  0.8&  1.7&  1.7\\
LeptonID &  1.2&  3.1&  0.9&  1.9\\
HadronID &  0.6&  0.6&  0.1&  0.1\\
$B \rightarrow D \ell \nu$ FF &  0.1&  0.1&  0.1&  0.1\\
$B \rightarrow D^* \ell \nu$ FF &  0.1&  0.2&  0.0&  0.0\\
$\mathcal{B}(B \rightarrow X_c \ell \nu)$ &  1.9&  1.9&  0.4&  0.3\\
Continuum normalization &  0.2&  0.2&  0.1&  0.1\\
Fake $D$ PDFs &  1.4&  1.5&  3.0&  2.8\\\hline
Total&  3.5&  4.6&  4.2&  4.4\\ \hline\hline
    \end{tabular}
    \end{adjustbox}
    \caption{Branching ratio results for the decays \bchdenu, \bchdmunu, \bneudenu, and \bneudmunu. The first uncertainty is statistical, and the second is systematic. The lower half of the table shows the various contributions to the systematic uncertainty, which are explained in more detail in Sect.~\hyperref[sec:syst]{4.3}} \label{tab:br}
\end{table}

\subsection{Partial width as a function of $w$}

We measure the partial widths $\Delta\Gamma_i=\Delta\mathcal{B}(B\to D\ell\nu)_i/\tau(B)$ in bin~$i$ based on the fit results. The results are shown in Table~\ref{tab:part}. 
We calculate bin-wise efficiencies as the ratios of reconstructed signal events in a given $w$ bin to generated events in that bin. By applying this efficiency correction we correct for overall acceptance effects, as well as migration of candidates to other bins due to the observed $w$ resolution (bin-by-bin unfolding).
\begin{table}[]
    \centering
    \begin{adjustbox}{center}
    \setlength{\tabcolsep}{5pt}
    \begin{tabular}{lllllll}
\hline \hline \\[-2.0ex]
& & & \multicolumn{4}{c}{$\Delta \Gamma_i / \Delta w [10^{-15}$ GeV]}\\[1.0ex]
$i$ & $w_{min}$ & $w_{max}$&  \bchdenu& \bchdmunu& \bneudenu& \bneudmunu\\[0.5ex] \hline\\[-2.0ex]
0 & 1.0 & 1.06& $0.80 \pm 0.40  \pm 0.60$& $0.75 \pm 0.39  \pm 0.70$& $1.31 \pm 1.16  \pm 0.23$& $0.81 \pm 1.01  \pm 0.21$\\
1 & 1.06 & 1.12& $4.55 \pm 0.49  \pm 0.68$& $4.59 \pm 0.48  \pm 0.77$& $2.22 \pm 1.10  \pm 0.11$& $5.23 \pm 1.08  \pm 0.27$\\
2 & 1.12 & 1.18& $7.60 \pm 0.53  \pm 0.73$& $7.55 \pm 0.51  \pm 0.88$& $5.11 \pm 1.03  \pm 0.23$& $5.30 \pm 0.99  \pm 0.23$\\
3 & 1.18 & 1.24& $11.15 \pm 0.53  \pm 0.71$& $10.08 \pm 0.52  \pm 0.92$& $7.63 \pm 0.99  \pm 0.33$& $9.24 \pm 0.93  \pm 0.41$\\
4 & 1.24 & 1.30& $12.37 \pm 0.53  \pm 0.67$& $14.37 \pm 0.52  \pm 1.01$& $13.75 \pm 0.83  \pm 0.55$& $13.73 \pm 0.85  \pm 0.58$\\
5 & 1.30 & 1.36& $17.94 \pm 0.54  \pm 0.76$& $16.30 \pm 0.53  \pm 1.01$& $17.58 \pm 0.81  \pm 0.69$& $17.27 \pm 0.79  \pm 0.74$\\
6 & 1.36 & 1.42& $19.93 \pm 0.56  \pm 0.76$& $19.46 \pm 0.55  \pm 1.06$& $20.19 \pm 0.77  \pm 0.80$& $21.25 \pm 0.76  \pm 0.93$\\
7 & 1.42 & 1.48& $22.95 \pm 0.60  \pm 0.79$& $23.25 \pm 0.61  \pm 1.05$& $23.10 \pm 0.85  \pm 0.94$& $24.18 \pm 0.82  \pm 1.06$\\
8 & 1.48 & 1.54& $23.98 \pm 0.83  \pm 0.87$& $26.75 \pm 1.18  \pm 1.02$& $26.48 \pm 1.01  \pm 1.11$& $23.85 \pm 1.11  \pm 1.06$\\
9 & 1.54 & 1.59& $26.73 \pm 1.25  \pm 1.10$& $27.78 \pm 1.58  \pm 1.21$& $29.58 \pm 1.34  \pm 1.36$& $30.64 \pm 1.65  \pm 1.47$\\[0.5ex]
\hline \hline
    \end{tabular}
        \end{adjustbox}

    \caption{Results for the partial widths, $\Gamma_i$, in the  charged and neutral channels. The first uncertainty is statistical, and the second is systematic. Refer to Sec. \hyperref[sec:syst]{4.3} for details on the systematic uncertainty.} \label{tab:part}
\end{table}

\subsection{Systematic uncertainties} \label{sec:syst}

\subsubsection{Systematic uncertainty on the branching fractions} \label{ssec:systbr}

The number of charged and neutral $B$ mesons in the data sample is calculated as
\begin{equation}
   N_{B^{\pm/0}} = 2  N_{B\bar B}  f_{+-/00}~\,,
\end{equation}
with \NBB~\cite{HFLAV:2019otj}, 
\begin{equation}
  f_{+-} = \frac{\Gamma\left(\Upsilon\left(4S\right) \to B^+ B^-\right)}{\Gamma\left(\Upsilon\left(4S\right)\right)_\mathrm{tot} } = 0.514 \pm 0.006~,
\end{equation}
and \cite{HFLAV:2019otj} 
\begin{equation}
  f_{00} = \frac{\Gamma\left(\Upsilon\left(4S\right) \to B^0 \overline{B}\right)}{\Gamma\left(\Upsilon\left(4S\right)\right)_\mathrm{tot} } = 0.486 \pm 0.006~.
\end{equation}
The uncertainties on $f$ and $N_{B \bar B}$ are added in quadrature to estimate the impact on the measured branching fraction.
To correct for mismodelling of the lepton-identification in the MC simulation compared to data, we apply momentum-and polar-angle-dependent corrections.

The difference between track-finding efficiencies in data and MC simulation is assigned as a systematic uncertainty per track, evaluated with a $e^+e^-\to \tau^+\tau^-$ control sample. A relative systematic uncertainty of $0.30\%$ is assigned for each of the final-state charged particles, resulting in $0.9\%$ and $1.2\%$ systematic uncertainties for $B^-$ and $B^0$ modes, respectively. The uncertainty on the branching fractions  $\BR(\Dz \to \Km \pip)$ = $(3.95\pm0.03)\%$ and $\BR(D^+ \to \Km \pip \pip)$ = $(9.38\pm 0.16)\%$~\cite{zyla:2020} also contributes a systematic uncertainty.

In independent studies of decays such as $J/\psi\to \ell^+\ell^-$ and $K^0_S \to \pi^+ \pi^-$ decays, correction factors are obtained for the reconstruction efficiency of leptons and the misidentification of hadrons as leptons. The lepton-identification correction factors are associated with uncertainties from statistical and systematic sources. By resampling the correction factors from Gaussian distributions while accounting for systematic error correlations, we generate 500 sets of correction values. The 500 sets are used to estimate the corresponding systematic uncertainty. A similar treatment is applied to correct hadron-identification MC mismodelling. Corrections are obtained from studies of the decay $D^{*+} \to D^0 (\to K^- \pi^+) \pi^+$. 

The form factors describe the effects of the strong interaction in the decay, which are parameterized as functions of $w = v_B \cdot v_{D^{(*)}}$. The impact of the uncertainty in form factors on the signal and \bdslnu PDFs has to be taken into account. To assess the model uncertainty of \bdlnu, we vary the form factor parameter $\rho^2$ in the parameterization of Caprini, Lellouch, and Neubert~\cite{Caprini:1997mu} by one standard deviation around its central values~\cite{HFLAV:2019otj}. For the \bdslnu model uncertainty, we restrict ourself to varying $\rho^2$ in the CLN paramterization as none of the selections applied in the analysis depend on the angles in the \bdslnu helicity frame.

To estimate the uncertainty arising from other $B \to X_c \ell \nu$ background effects, we vary the $B \to D^* \ell \nu$ and $B \to D^{**} \ell \nu$ branching fractions within their uncertainties. Additionally, the gap between inclusive $B \to X_c \ell \nu$ measurements and the sum of exclusive measurements, is accounted for in MC generation by the hypothetical gap modes $B \to D \eta \ell \nu$ and $B \to D^* \eta \ell \nu$. Because they have not been measured, we assign 100\% uncertainty to the gap mode branching fractions.

We determine the continuum normalization by scaling off-resonance data to on-resonance luminosity. The normalization uncertainty is due to the limited off-resonance sample size.

The distribution of the fake $D$ component in \cosby~is adjusted to match the shape obtained in the $m_D < 1.85$~GeV sideband where fake $D$ background is dominant. To estimate the uncertainty of this reshaping, we assign a 100\% uncertainty on the weights applied to correct the fake $D$ shape. 
\subsubsection{Systematic uncertainty on the partial widths}
\begin{table}[!h]
    \footnotesize
    
    \centering
    \setlength{\tabcolsep}{3pt}
    \begin{tabular}{lrrrrrrrrrr}
\hline \hline \\[-3.0ex]

\multicolumn{1}{l}{\bchdenu}& \multicolumn{10}{c}{$\Delta\Gamma_{i} / \Delta w$ uncertainty $[\%]$}\\[1.0ex]
& 0& 1& 2& 3& 4& 5& 6& 7& 8& 9\\[0.5ex]\hline\\[-2.0ex]
$N_{BB}$ and $f_{+-}/f_{00}$ &  1.9&  1.9&  1.9&  1.9&  1.9&  1.9&  1.9&  1.9&  1.9&  1.9\\
Tracking efficiency &  0.9&  0.9&  0.9&  0.9&  0.9&  0.9&  0.9&  0.9&  0.9&  0.9\\
$\mathcal{B}(D \rightarrow K \pi (\pi))$ &  0.8&  0.8&  0.8&  0.8&  0.8&  0.8&  0.8&  0.8&  0.8&  0.8\\
LeptonID&  15.5&  3.8&  2.3&  1.6&  1.2&  1.0&  1.0&  1.0&  1.2&  1.5\\
HadronID&  5.2&  2.9&  2.1&  1.3&  1.0&  0.6&  0.5&  0.3&  0.3&  0.4\\
$B \rightarrow D \ell \nu$ FF&  2.6&  2.2&  1.8&  1.4&  1.0&  0.6&  0.1&  0.3&  0.8&  1.3\\
$B \rightarrow D^* \ell \nu$ FF&  34.3&  4.7&  1.9&  0.5&  0.1&  0.4&  0.4&  0.3&  0.0&  0.2\\
$\mathcal{B}(B \rightarrow X_c \ell \nu)$&  40.6&  9.7&  6.6&  4.1&  3.3&  2.0&  1.4&  0.6&  0.2&  0.7\\
Continuum normalization&  0.2&  0.1&  0.0&  0.0&  0.0&  0.0&  0.0&  0.1&  0.8&  1.7\\
Fake $D$ PDFs&  15.4&  12.7&  8.8&  5.3&  5.0&  2.5&  2.3&  1.0&  1.1&  0.7\\
$\tau_{B}$ &  0.2&  0.2&  0.2&  0.2&  0.2&  0.2&  0.2&  0.2&  0.2&  0.2\\ \hline
Total&  70.6&  17.9&  11.8&  7.5&  6.6&  4.2&  3.7&  2.8&  3.0&  3.4\\[0.5ex]
\hline \hline\\[0.5ex]
    \end{tabular}
    \begin{tabular}{lrrrrrrrrrr}
\hline \hline \\[-3.0ex]

\multicolumn{1}{l}{\bchdmunu}& \multicolumn{10}{c}{$\Delta\Gamma_{i} / \Delta w$ uncertainty $[\%]$}\\[1.0ex]
& 0& 1& 2& 3& 4& 5& 6& 7& 8& 9\\[0.5ex]\hline\\[-2.0ex]
$N_{BB}$ and $f_{+-}/f_{00}$ &  1.9&  1.9&  1.9&  1.9&  1.9&  1.9&  1.9&  1.9&  1.9&  1.9\\
Tracking efficiency &  0.9&  0.9&  0.9&  0.9&  0.9&  0.9&  0.9&  0.9&  0.9&  0.9\\
$\mathcal{B}(D \rightarrow K \pi (\pi))$ &  0.8&  0.8&  0.8&  0.8&  0.8&  0.8&  0.8&  0.8&  0.8&  0.8\\
LeptonID&  19.8&  7.6&  6.6&  6.0&  4.9&  4.4&  3.9&  2.9&  1.5&  2.2\\
HadronID&  5.6&  2.8&  2.3&  1.5&  0.9&  0.7&  0.5&  0.3&  0.2&  0.3\\
$B \rightarrow D \ell \nu$ FF&  2.7&  2.3&  1.9&  1.5&  1.0&  0.6&  0.1&  0.3&  0.9&  1.3\\
$B \rightarrow D^* \ell \nu$ FF&  37.5&  4.8&  2.0&  0.6&  0.1&  0.4&  0.5&  0.3&  0.3&  0.3\\
$\mathcal{B}(B \rightarrow X_c \ell \nu)$&  46.3&  10.0&  6.8&  4.7&  2.9&  2.1&  1.5&  0.7&  0.8&  0.7\\
Continuum normalization&  0.8&  0.1&  0.0&  0.0&  0.0&  0.0&  0.0&  0.1&  0.8&  1.5\\
Fake $D$ PDFs&  19.3&  12.4&  8.9&  6.3&  3.7&  3.3&  2.1&  1.2&  0.7&  0.8\\
$\tau_{B}$ &  0.2&  0.2&  0.2&  0.2&  0.2&  0.2&  0.2&  0.2&  0.2&  0.2\\ \hline
Total&  91.9&  20.2&  14.1&  10.5&  7.3&  6.3&  5.1&  3.9&  3.1&  3.9\\[0.5ex]
\hline \hline
    \end{tabular}
    \caption{Systematic uncertainties on $\Delta\Gamma / \Delta w$ in $w$ bins for the $B^+$ modes.  The total uncertainties correspond to the sum of the systematic error components taking into account correlated uncertainties.}
    \label{tab:uncertainties_charged}
\end{table}
\begin{table}[]
    \footnotesize

    \setlength{\tabcolsep}{3pt}

    \centering
    \begin{tabular}{lrrrrrrrrrr}
\hline \hline \\[-3.0ex]
\multicolumn{1}{l}{\bneudenu }& \multicolumn{10}{c}{$\Delta\Gamma_{i} / \Delta w$ uncertainty $[\%]$}\\[1.0ex]
& 0& 1& 2& 3& 4& 5& 6& 7& 8& 9\\[0.5ex]\hline\\[-2.0ex]
$N_{BB}$ and $f_{+-}/f_{00}$ &  1.9&  1.9&  1.9&  1.9&  1.9&  1.9&  1.9&  1.9&  1.9&  1.9\\
Tracking efficiency &  1.2&  1.2&  1.2&  1.2&  1.2&  1.2&  1.2&  1.2&  1.2&  1.2\\
$\mathcal{B}(D \rightarrow K \pi (\pi))$ &  1.7&  1.7&  1.7&  1.7&  1.7&  1.7&  1.7&  1.7&  1.7&  1.7\\
LeptonID&  1.8&  0.6&  0.6&  0.7&  0.5&  0.6&  0.8&  1.0&  1.1&  1.7\\
HadronID&  1.0&  0.8&  0.2&  0.2&  0.2&  0.2&  0.2&  0.1&  0.2&  0.2\\
$B \rightarrow D \ell \nu$ FF&  2.6&  2.2&  1.8&  1.4&  1.0&  0.6&  0.1&  0.4&  0.8&  1.3\\
$B \rightarrow D^* \ell \nu$ FF&  0.2&  0.0&  0.0&  0.0&  0.0&  0.0&  0.0&  0.0&  0.0&  0.0\\
$\mathcal{B}(B \rightarrow X_c \ell \nu)$&  13.7&  2.7&  1.4&  1.5&  0.5&  0.3&  0.3&  0.3&  0.5&  0.4\\
Continuum normalization&  0.4&  0.1&  0.0&  0.0&  0.0&  0.0&  0.1&  0.1&  0.4&  0.7\\
Fake $D$ PDFs&  32.0&  20.1&  10.1&  6.9&  2.9&  2.2&  1.0&  2.6&  2.1&  1.4\\
$\tau_{B}$ &  0.3&  0.3&  0.3&  0.3&  0.3&  0.3&  0.3&  0.3&  0.3&  0.3\\ \hline
Total&  37.9&  21.4&  10.7&  8.0&  4.3&  3.7&  3.1&  4.0&  3.9&  4.0\\[0.5ex]
\hline \hline\\[0.5ex]
    \end{tabular}
    \begin{tabular}{lrrrrrrrrrr}
\hline \hline \\[-3.0ex]

\multicolumn{1}{l}{\bneudmunu }& \multicolumn{10}{c}{$\Delta\Gamma_{i} / \Delta w$ uncertainty $[\%]$}\\[1.0ex]
& 0& 1& 2& 3& 4& 5& 6& 7& 8& 9\\[0.5ex]\hline\\[-2.0ex]
$N_{BB}$ and $f_{+-}/f_{00}$ &  1.9&  1.9&  1.9&  1.9&  1.9&  1.9&  1.9&  1.9&  1.9&  1.9\\
Tracking efficiency &  1.2&  1.2&  1.2&  1.2&  1.2&  1.2&  1.2&  1.2&  1.2&  1.2\\
$\mathcal{B}(D \rightarrow K \pi (\pi))$ &  1.7&  1.7&  1.7&  1.7&  1.7&  1.7&  1.7&  1.7&  1.7&  1.7\\
LeptonID&  18.3&  2.7&  1.4&  1.9&  1.9&  2.0&  2.2&  2.1&  1.8&  2.2\\
HadronID&  2.1&  0.4&  0.3&  0.2&  0.2&  0.2&  0.1&  0.1&  0.2&  0.2\\
$B \rightarrow D \ell \nu$ FF&  2.5&  2.1&  1.7&  1.3&  0.9&  0.5&  0.1&  0.3&  0.8&  1.2\\
$B \rightarrow D^* \ell \nu$ FF&  0.4&  0.0&  0.0&  0.0&  0.0&  0.0&  0.0&  0.0&  0.0&  0.0\\
$\mathcal{B}(B \rightarrow X_c \ell \nu)$&  20.2&  1.8&  0.8&  1.0&  0.3&  0.2&  0.3&  0.3&  1.0&  0.3\\
Continuum normalization&  0.8&  0.1&  0.1&  0.0&  0.0&  0.0&  0.0&  0.1&  0.4&  0.7\\
Fake $D$ PDFs&  32.8&  10.4&  10.1&  5.2&  2.4&  2.0&  1.1&  2.5&  2.2&  2.3\\
$\tau_{B}$ &  0.3&  0.3&  0.3&  0.3&  0.3&  0.3&  0.3&  0.3&  0.3&  0.3\\ \hline
Total&  44.1&  11.6&  11.2&  6.7&  4.3&  4.0&  3.8&  4.3&  4.3&  4.6\\[0.5ex]
\hline \hline
    \end{tabular}
    \caption{Systematic uncertainties on $\Delta\Gamma / \Delta w$ in $w$ bins for the $B^0$ modes. The total uncertainties correspond to the sum of the systematic error components taking into account correlated uncertainties. }
    \label{tab:uncertainties_neutral}
\end{table}
To estimate systematic uncertainties on the values of $\Delta \Gamma_{i} / \Delta w$ and their correlations, we use toy MC (or fast MC pseudo-experiments). For each systematic uncertainty listed in Sect.~\hyperref[ssec:systbr]{4.3.1}  we generate an altered MC sample by varying the corresponding values within their uncertainties. The fit for $\Delta \Gamma_{i} / \Delta w$ is repeated in each $w$ bin for each altered MC. The width of the distribution of $\Delta \Gamma_{i} / \Delta w$ results for 500 altered samples is taken as systematic uncertainty for the corresponding effect. The systematic uncertainties on  $\Delta \Gamma_{i} / \Delta w$ are listed in Tables \ref{tab:uncertainties_charged} and \ref{tab:uncertainties_neutral}.

Correlation matrices are then obtained as  
\begin{equation}
    { \rho_{i j}=\frac{\left\langle\left(\frac{\Delta \Gamma_{i}}{\Delta w}-\left\langle\frac{\Delta \Gamma_{i}}{\Delta w}\right\rangle\right)\left(\frac{\Delta \Gamma_{j}}{\Delta w}-\left\langle\frac{\Delta \Gamma_{j}}{\Delta w}\right\rangle\right)\right\rangle}{\sqrt{\left\langle\left(\frac{\Delta \Gamma_{i}}{\Delta w}-\left\langle\frac{\Delta \Gamma_{i}}{\Delta w}\right\rangle\right)^{2}\right\rangle} \sqrt{\left\langle\left(\frac{\Delta \Gamma_{j}}{\Delta w}-\left\langle\frac{\Delta \Gamma_{j}}{\Delta w}\right\rangle\right)^{2}\right\rangle}}}.
\end{equation}
Angle brackets denote averaging over the generated toy MC samples. Statistical uncertainties are treated as uncorrelated.

The covariance matrix is calculated as $C_{ij} = \rho_{ij} \sigma_i \sigma_j,$
where $\sigma_i$ is the standard deviation of $\Delta \Gamma_{i} / \Delta w$ in bin $i$.

In addition to the systematic uncertainties listed in Sect.~\hyperref[ssec:systbr]{4.3.1}, there is an uncertainty associated with $B$ meson lifetimes $\tau_{B^{\pm}} = 1.638 \pm 0.004$ and $\tau_{B^{0}} = 1.519 \pm 0.004$, which are inputs in the partial-width determination.

\section{Determination of $|V_{cb}|$} \label{sec:vcb}
\begin{table}[]
    \centering
    \begin{adjustbox}{center}
    \setlength{\tabcolsep}{10pt}

    \begin{tabular}{lrrrr}
    \hline \hline\\[-2.0ex]
&  \bchdenu& \bchdmunu& \bneudenu& \bneudmunu\\[0.5ex]\hline \\[-2.0ex]
$a_{+,0} \times 10^{2}$ & $ 1.27 \pm  0.01$& $ 1.27 \pm  0.01$& $ 1.26 \pm  0.01$& $ 1.26 \pm  0.01$\\
$a_{+,1} \times 10$ & $-0.96 \pm  0.03$& $-0.95 \pm  0.03$& $-0.95 \pm  0.03$& $-0.95 \pm  0.03$\\
$a_{+,2}$ & $ 0.44 \pm  0.17$& $ 0.33 \pm  0.17$& $ 0.36 \pm  0.17$& $ 0.41 \pm  0.17$\\
$a_{+,3}$ & $-3.06 \pm  2.76$& $ 0.73 \pm  3.09$& $ 0.10 \pm  2.79$& $-1.41 \pm  2.93$\\
$a_{0,1} \times 10$ & $-0.59 \pm  0.03$& $-0.58 \pm  0.03$& $-0.58 \pm  0.03$& $-0.59 \pm  0.03$\\
$a_{0,2}$ & $ 0.29 \pm  0.15$& $ 0.18 \pm  0.15$& $ 0.21 \pm  0.15$& $ 0.25 \pm  0.15$\\
$a_{0,3}$ & $-3.59 \pm  2.73$& $ 0.35 \pm  3.09$& $-0.13 \pm  2.75$& $-1.76 \pm  2.89$\\[0.5ex] \hline\\[-2.0ex]
$\eta_{EW} |V_{cb}| \times 10^3$  & $ 38.72 \pm  1.09$& $ 37.91 \pm  1.27$& $ 38.47 \pm  1.10$& $ 38.89 \pm  1.17$\\
$\chi^2 / ndf$& $18.2/14$& $12.3/14$& $11.0/14$& $15.1/14$\\[0.5ex] \hline\hline\\[0.3ex]
    \end{tabular}
    \end{adjustbox}
    \caption{Result of the BGL fit to the partial widths $\Delta\Gamma_i/\Delta w$ (Table~\ref{tab:part}) and to two lattice QCD calculations of the form factors $f_+$ and $f_0$~\cite{Lattice:2015rga,Na:2015kha}. The BGL series is truncated at $N=3$.} \label{tab:bgl}
\end{table}
\begin{figure}
    \centering
    \includegraphics[width=.47\columnwidth]{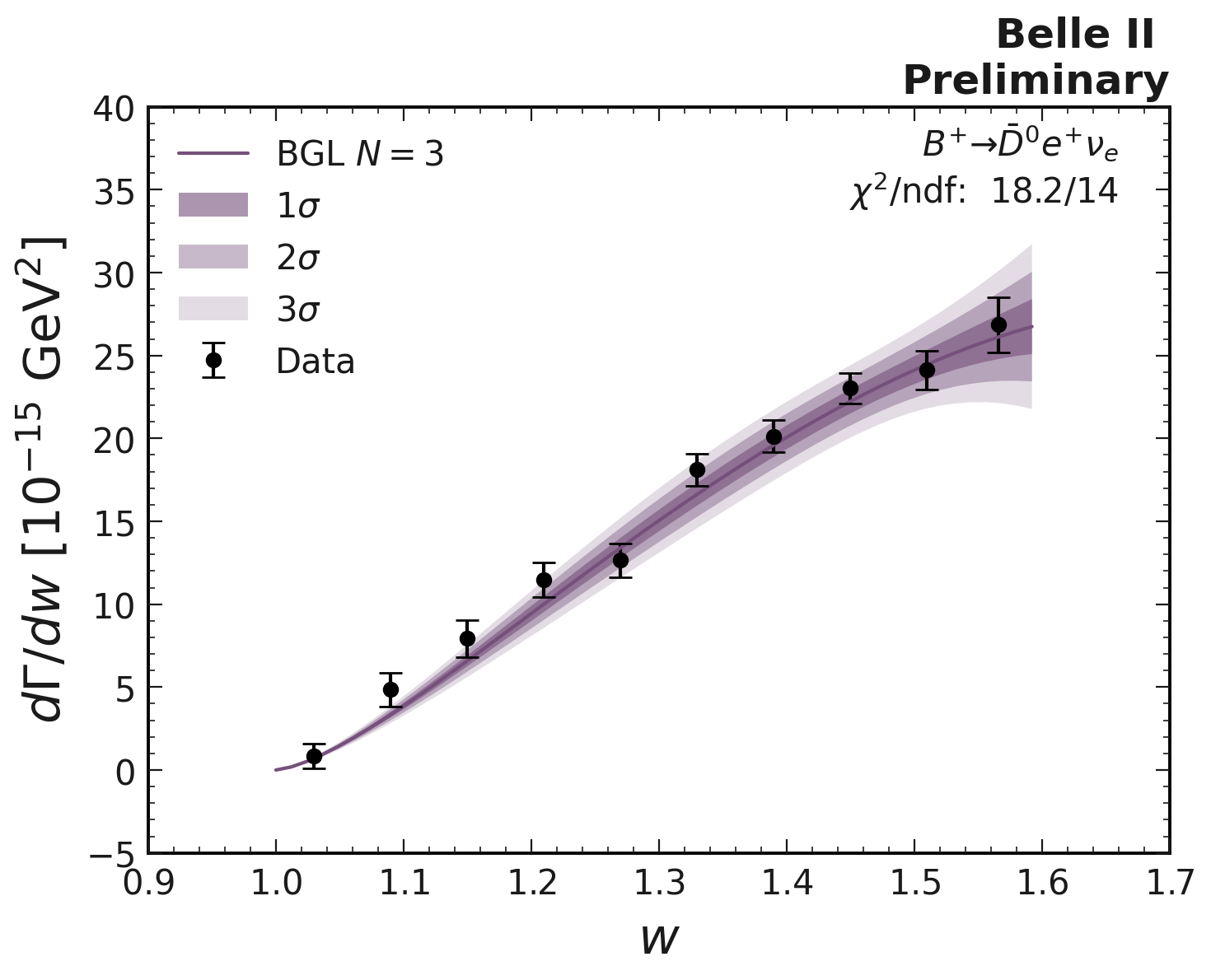}
    \includegraphics[width=.47\columnwidth]{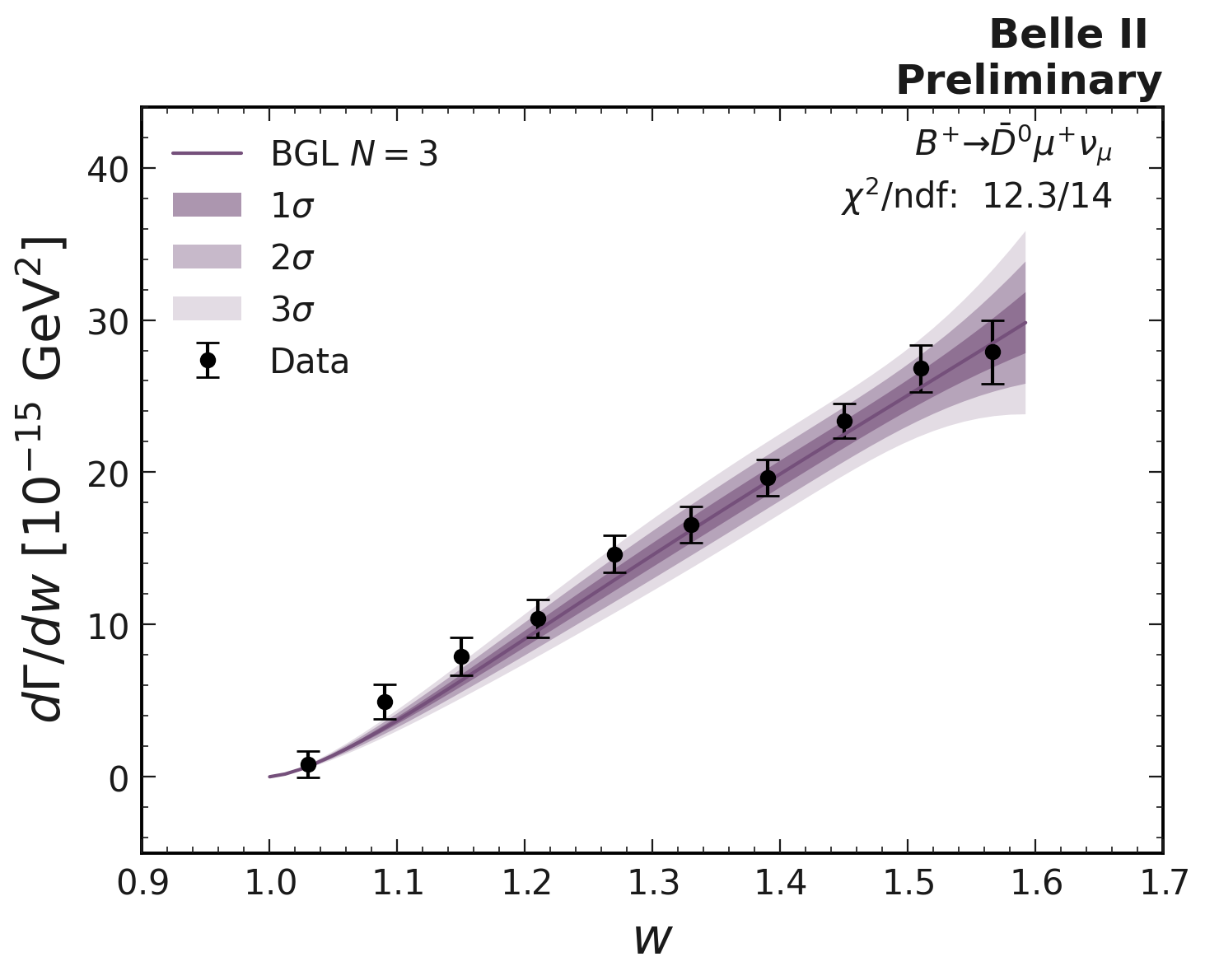}
    
    \includegraphics[width=.47\columnwidth]{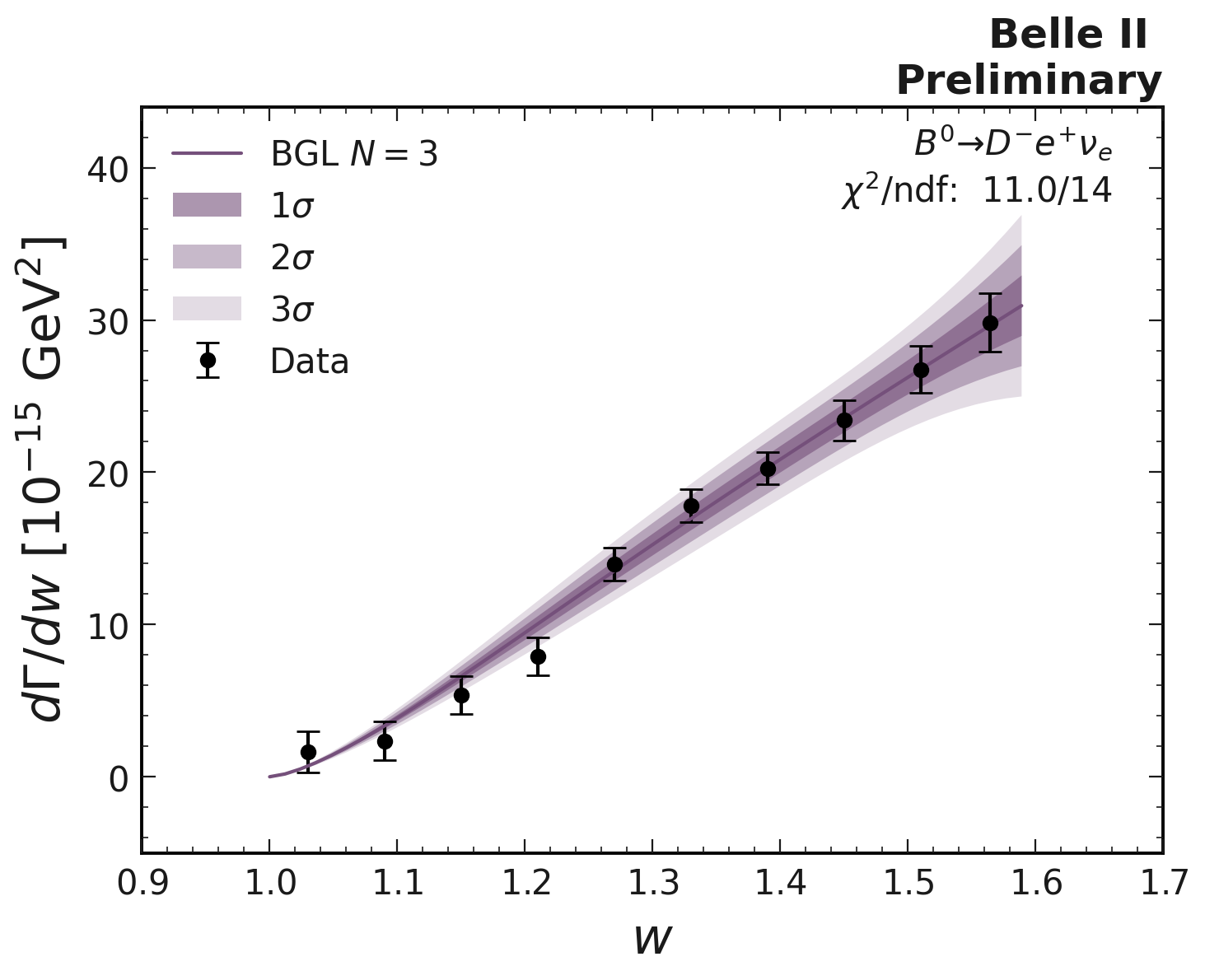}
    \includegraphics[width=.47\columnwidth]{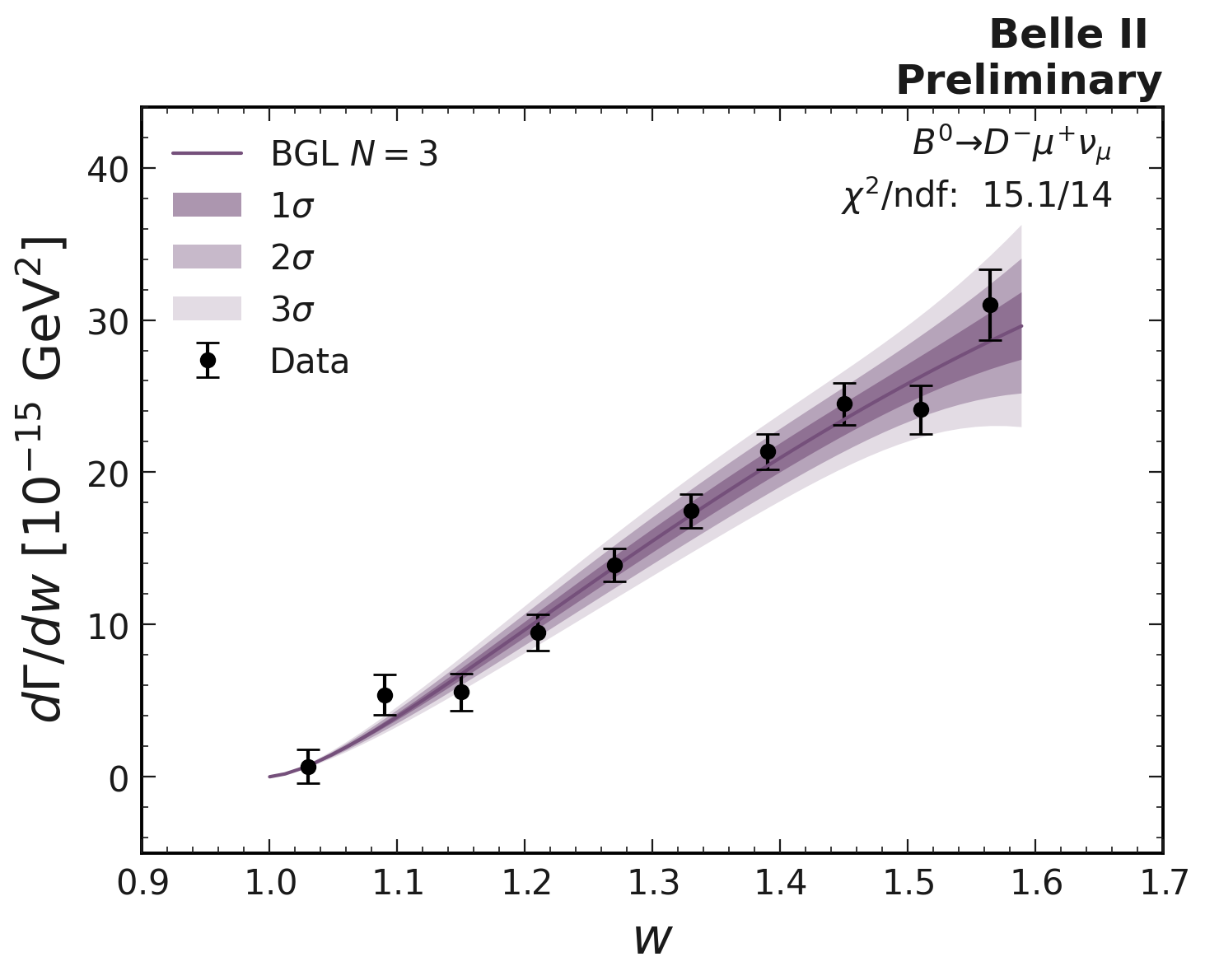}
    \caption{Observed $B\to D\ell\nu$ differential decay rates as functions of $w$ with results of a combined fit to experimental and lattice QCD (FNAL/MILC and HPQCD) data. The fit results are shown separately for the charged and neutral channels and the electron and muon samples.}
    \label{fig:rate}
\end{figure}

\begin{figure}
    \centering
    \includegraphics[width=.47\columnwidth]{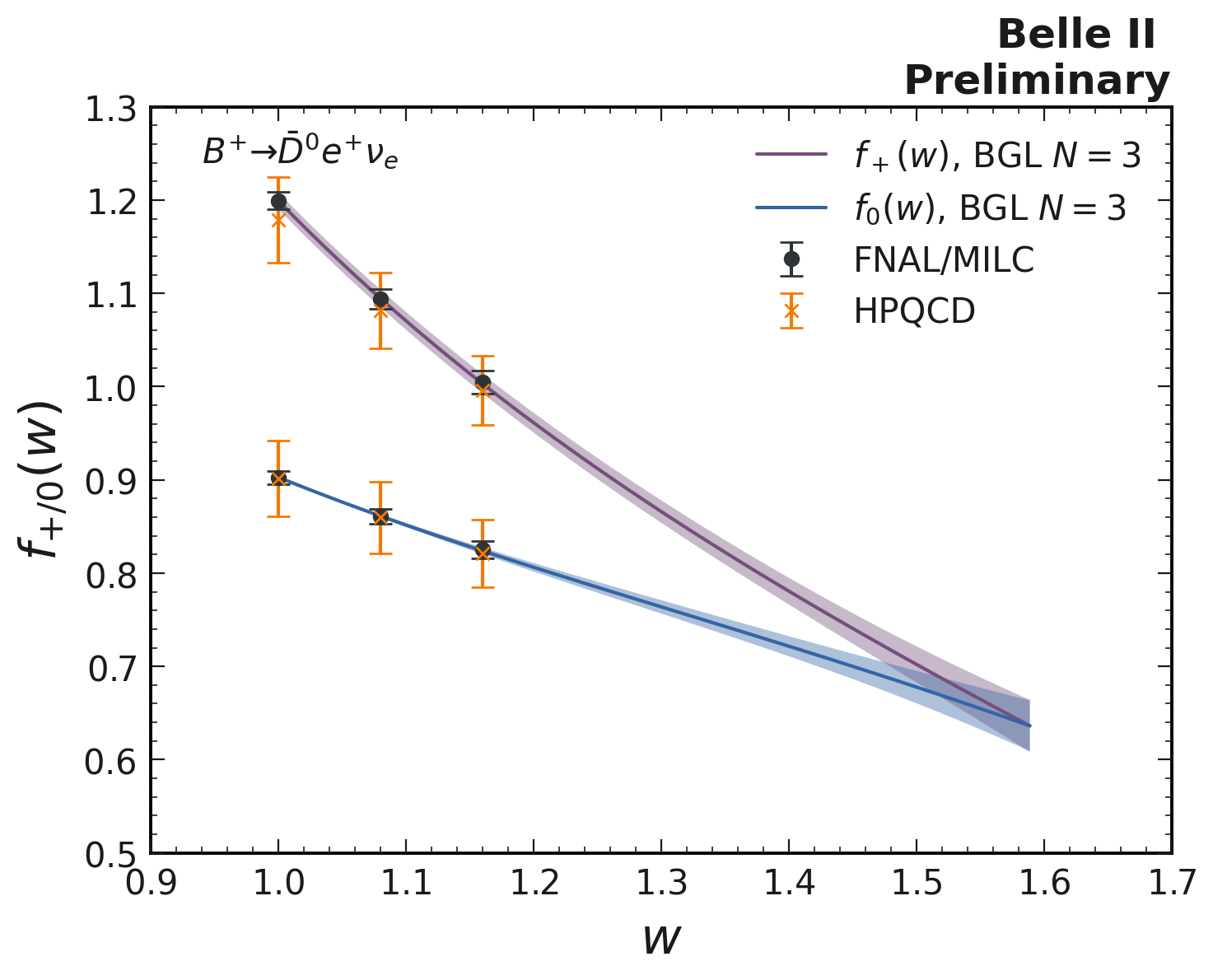}
    \includegraphics[width=.47\columnwidth]{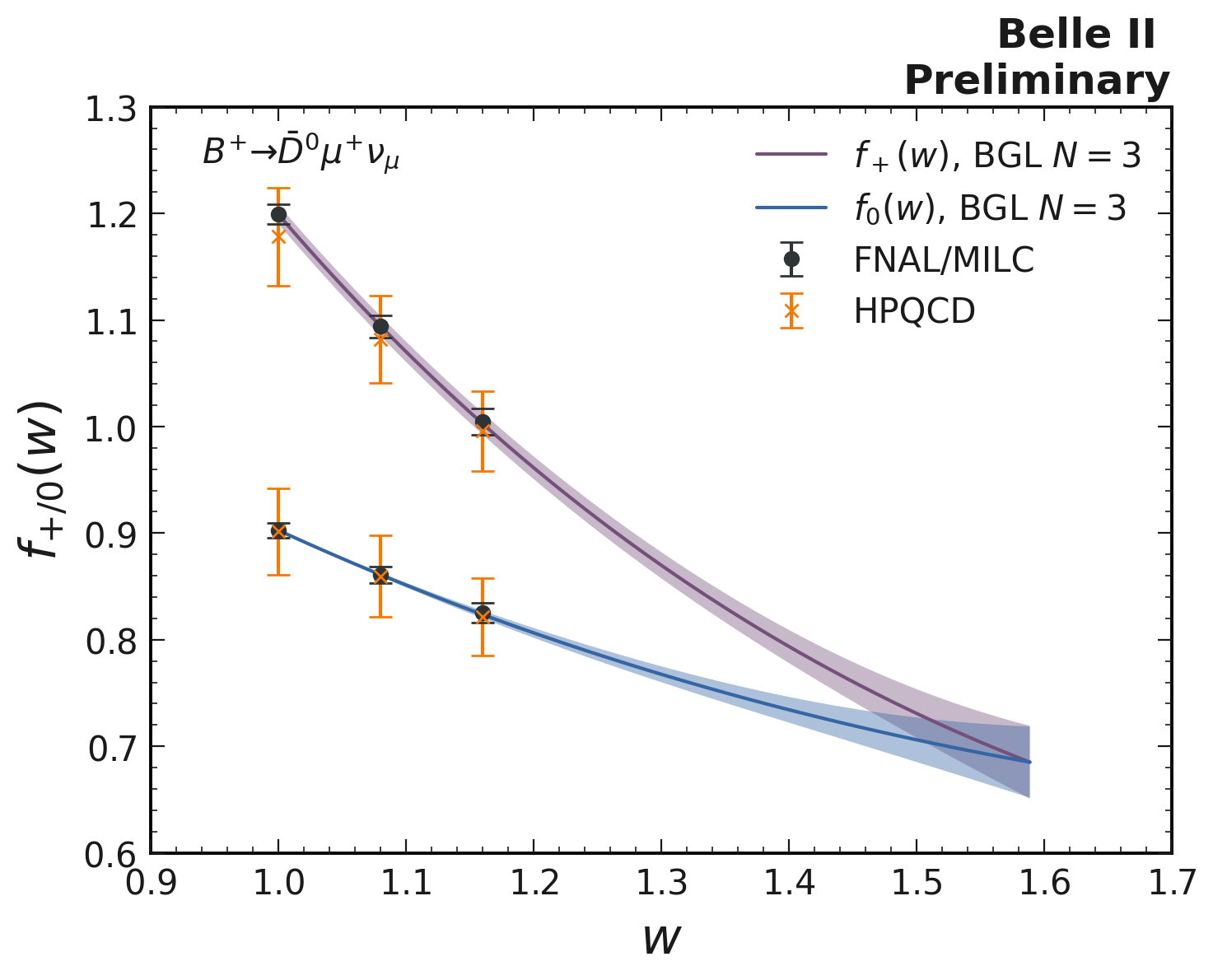}
    
    \includegraphics[width=.47\columnwidth]{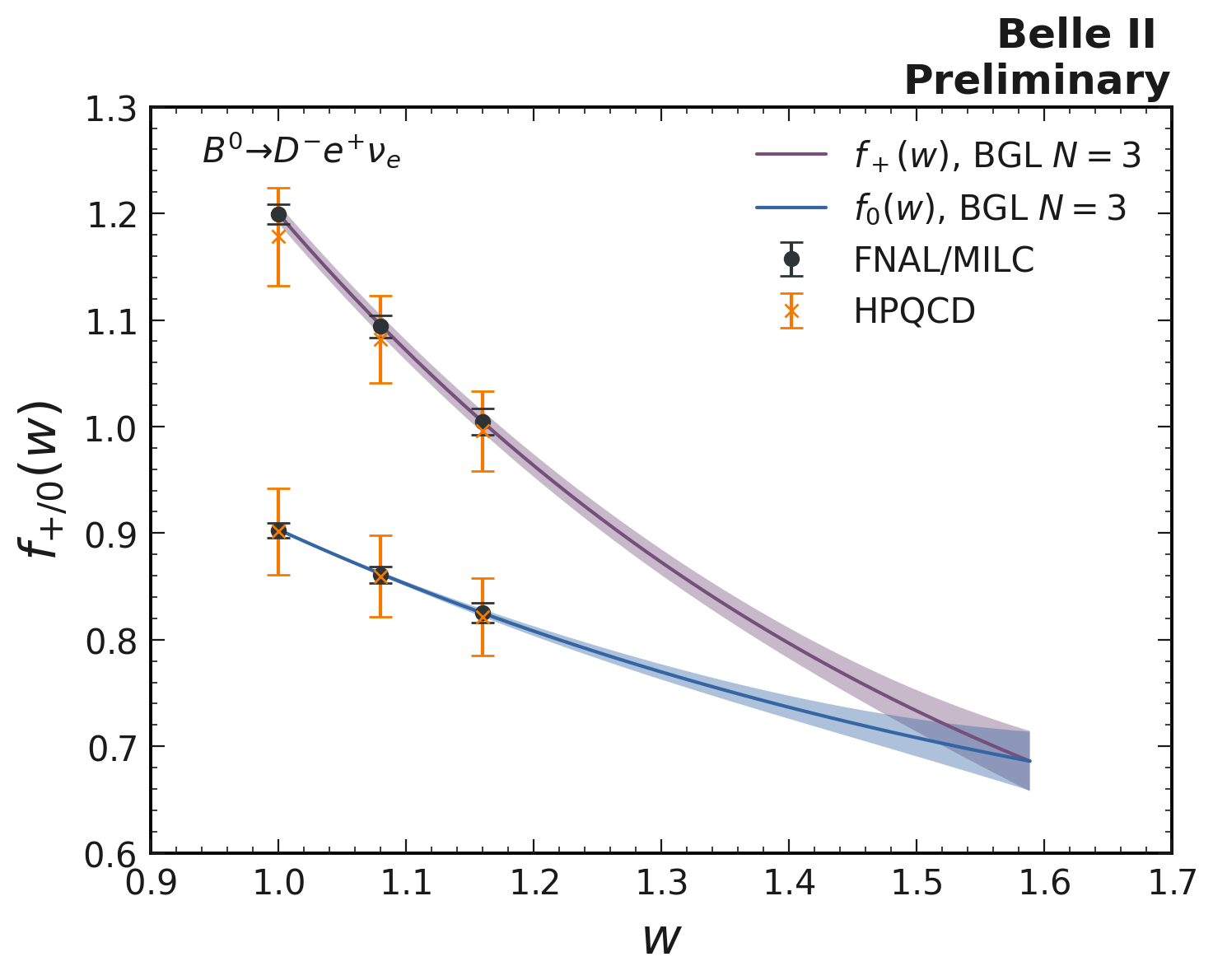}
    \includegraphics[width=.47\columnwidth]{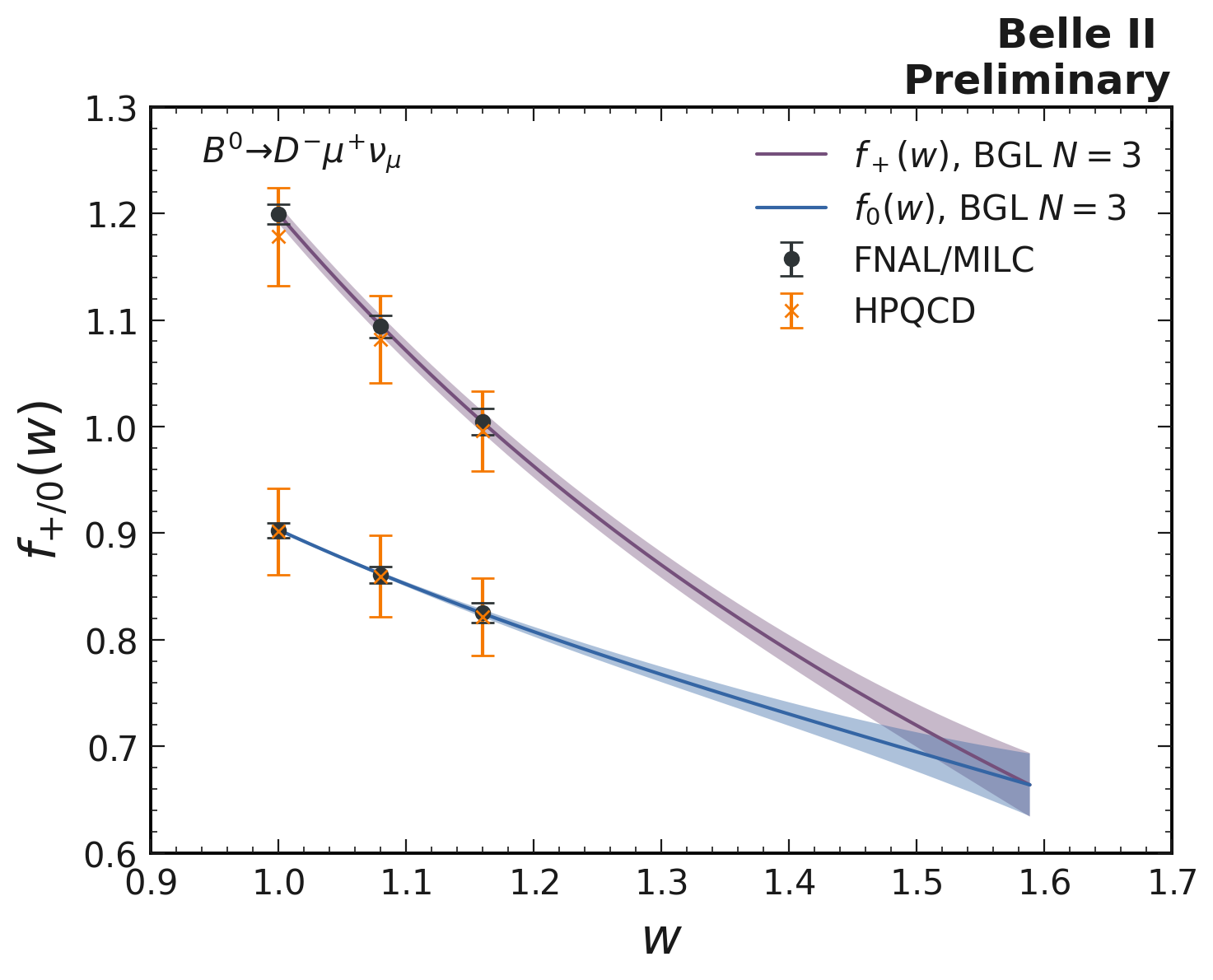}
    \caption{Form factors of the decay $B\to D\ell\nu$ as functions of $w$ and result of the combined fit to experimental and lattice QCD (FNAL/MILC and HPQCD) data. The fit results are shown separately for the charged and neutral channels and the electron and muon samples.}
    \label{fig:ff}
\end{figure}

To determine the CKM parameter $|V_{cb}|$, we perform a combined $\chi^2$~fit to the BGL form factor (Eq.~\ref{eq:BGL}) and lattice QCD calculations of the $f_+$ and $f_0$ form factors at $w>1$, by minimizing 
\begin{eqnarray}
  \chi^2 & = & \sum\limits_{i,j}\left(\dGidwf-\dGiBGLdwf\right)\mathbf{C}^{-1}_{ij}\left(\dGjdwf-\dGjBGLdwf\right) 
  +\\
  \lefteqn{+\sum\limits_{k,l}\left(f^\mathrm{LQCD}_{+,0}\left(w_k\right)-f^\mathrm{BGL}_{+,0}\left(w_k\right)\right)\mathbf{D}^{-1}_{kl}
  \left(f^\mathrm{LQCD}_{+,0} \left(w_l\right)-f^\mathrm {BGL}_{+,0}\left(w_l\right)\right).} \nonumber~
\end{eqnarray}
Here, the {\dGidw} values are taken from Table~\ref{tab:part} and {\dGiBGLdw} are the partial widths calculated using Eqs.~\ref{eq:rate}, 
\ref{eq:ff}, \ref{eq:BGL} and \ref{eq:kinematic}. The covariance matrix $\mathbf{C}$ includes the statistical and systematic uncertainties in the measurements of $\Delta\Gamma_i/\Delta w$. The data are fit together with predictions of lattice QCD (LQCD), which are available for the form factors $f_+(w)$ and $f_0(w)$ at select $w$ values. The second sum runs over all LQCD predictions included in the fit and the corresponding covariance matrix $\mathbf{D}$ contains the LQCD uncertainty in these predictions. We use lattice data obtained by the FNAL/MILC and HPQCD collaborations~\cite{Lattice:2015rga,Na:2015kha}. 
Both LQCD calculations are dominated by their systematic uncertainties. Both lattice calculations are sufficiently independent to allow their use in the same fit~\cite{https://doi.org/10.48550/arxiv.2111.09849}.

LQCD yields results for both the $f_+$ and $f_0$ form factors while the experimental distribution $\Delta\Gamma_i/\Delta w$ 
depends on $f_+$ only. Using the kinematic constraint from Eq.~\ref{eq:kinematic}, we include the LQCD results for $f_0$ in the fit, allowing us 
to better constrain $f_+$. Following Ref.~\cite{Lattice:2015rga}, we implement this constraint by expressing $a_{0,0}$ in terms of the other $a_{+,n}$ 
and $a_{0,n}$ coefficients. FNAL/MILC obtains values for both the $f_+$ and the $f_0$ form factors at $w$ values of 1, 1.08, and 1.16. The full covariance matrix for these six measurements is available in 
Table~VII of Ref.~\cite{Lattice:2015rga}. The form factors determined by HPQCD~\cite{Na:2015kha} are presented as fit results in the Bourrely, Caprini, and Lellouch parameterization~\cite{Bourrely:2008za} and have been transformed into extrapolations for $f_+$ and $f_0$ at $w=1, 1.08,$ and $1.16$ in Ref.~\cite{Belle:2015pkj}. We use these form factor results for the fit described in this section.

Following Ref.~\cite{Belle:2015pkj} we truncate the BGL~series at $N=3$ and obtain the fit result shown in Table~\ref{tab:bgl} and Figs.~\ref{fig:rate} and \ref{fig:ff}. A weighted average over the four samples ($B^- \to D^{0} e^{-} \overline{\nu}_e$, $B^- \to D^{0} \mu^{-} \overline{\nu}_\mu$, $B^0 \to D^{-} e^{+} \nu_e$, $B^0 \to D^{-} \mu^{+} \nu_\mu$) conservatively assuming full correlation of uncertainties yields \resVcb.

\section{Summary}

We reconstruct the decays $B\to D\ell\nu$ ($B^0\to D^-\ell^+\nu$ and $B^+\to\bar D^0\ell^+\nu$~\cite{cc}) in $\Upsilon(4S)$~events and perform a determination of the CKM parameter $|V_{cb}|$ using a Belle II data sample corresponding to \lumi. We extract the partial decay rates in ten bins of $w$ and perform a fit to the BGL expression of the form factor~\cite{Boyd:1994tt} and to two lattice QCD calculations~\cite{Lattice:2015rga,Na:2015kha}. The result in terms of $\eta_\mathrm{EW}|V_{cb}|$ is shown in Table~\ref{tab:bgl}, where $\eta_\mathrm{EW}$ is a small electroweak correction. The weighted average over the four samples (\bchdenu, \bchdmunu, \bneudenu, and \bneudmunu) assuming full correlation of uncertainties yields
\begin{equation}
  \resVcb ,
\end{equation}
in agreement with current world average estimates. The error quoted for $|V_{cb}|$ includes experimental and theoretical uncertainties. Assuming $\eta_\mathrm{EW}=1.0066\pm 0.0050$~\cite{Sirlin:1981ie}, we finally obtain \resVcbnoeta.    

\section{Acknowledgments}
We thank the SuperKEKB group for the excellent operation of the accelerator; the KEK cryogenics group for the efficient operation of the solenoid and the KEK computer group for on-site computing support.
\bibliographystyle{apsrev4-2} 
\bibliography{conf.bib}

\begin{thebibliography}{27}
\bibitem [{cc()}]{cc}%
  \BibitemOpen
  \href@noop {} {}\bibinfo {note} {Charged conjugation is implied throughout
  this document.}\BibitemShut {Stop}%
\end{thebibliography}
\end{document}